\documentstyle[twocolumn,eqsecnum,aps]{revtex}

\begin{document}
\draft
\preprint{}
\title{ Microscopic spin interactions in the CMR manganites}
\author{J. A. Fernandez-Baca,$^1$ Pengcheng Dai,$^{2,1}$
H. Kawano-Furukawa,$^3$ H. Yoshizawa,$^{4}$ E. W. Plummer, $^{2,1}$ S. Katano,$^{5}$
Y. Tomioka,$^6$ and Y. Tokura$^{6,7}$}
\address{$^1$Solid State Division, Oak Ridge National Laboratory, Oak Ridge, Tennessee 37831-6393}
\address{$^2$ Department of Physics and Astronomy, University of Tennessee, Knoxville, Tennessee 37966}
\address{$^3$ Dept. of Physics, Faculty of Science, Ochanomizu University
Ootsuka 2-1-1, Bunkyo-ku, Tokyo 112-8610, Japan}
\address{$^4$ Neutron Scattering Laboratory, Institute for Solid State Physics, University of Tokyo, Shirakata 106-1, Tokai, Ibaraki 319-1106, Japan}
\address{$^5$Advanced Science Research Center, Japan Atomic Energy Research Institute (JAERI), Tokai, Ibaraki  319-1195, Japan}
\address{$^6$Joint Research Center for Atom Technology (JRCAT), Tsukuba
305-8562, Japan}
\address{$^7$Department of Applied Physics, University of Tokyo, Tokyo
113-8656, Japan}
\date{\today}
\maketitle
\begin{abstract}
Using inelastic neutron scattering we measured the microscopic magnetic coupling associated with the ferromagnetic (FM) clusters of the ``colossal magnetoresistance'' (CMR) compound Pr$_{0.70}$Ca$_{0.30}$MnO$_3$.  When the insulating to metal (I-M) transition is induced by an external magnetic field there is a discontinuous change in the spin-wave stiffness constant. This result indicates that the probed regions undergo a first-order transition from an insulating to a metallic state, and that there are no FM metallic domains in the insulating region. We argue that the I-M transition in the CMR manganites is more complex than the simple percolation of large FM metallic clusters.
 
\end{abstract}


\narrowtext

The understanding of the colossal magnetoresistance (CMR) effect -- the unusually large change in electrical resistance in response to a magnetic field --  in certain materials is among the most interesting unresolved problems in condensed matter physics\cite{imada}. 
The most studied of these materials are the doped perovskite manganites  
$A_{1-x}B_x$MnO$_3$ [where $A$ is a trivalent ion 
(La$^{3+}$, Pr$^{3+}$, Nd$^{3+}$, etc) and $B$ is a divalent ion ( Ca$^{2+}$ or Sr$^{2+}$)] 
with $x\approx 0.3$.
The basic microscopic mechanism responsible for the CMR effect is believed to be   
the double-exchange (DE) interaction \cite{zener}, where 
ferromagnetism and electrical conductivity arise from hopping of the 
itinerant $e_g$ electrons from trivalent Mn$^{3+}$ to tetravalent Mn$^{4+}$ sites. 
The physics of the CMR effect, however, is far from being completely understood. 
For example, recent calculations suggest that these materials are intrinsically inhomogeneous, and have a strong tendency to spatial electronic phase separation \cite{dagotto}.   
Experimentally, the existence of mesoscopic phase separation, in which the I-M transition is achieved via the percolation of large metallic clusters,
has been suggested by electron microscopy and tunneling experiments \cite{uehara,fath}. 
If this mesoscopic phase 
separation scenario is central to the CMR physics, it should be universal and occur in all CMR manganites.  
In this Letter, we show that the I-M transition in Pr$_{0.70}$Ca$_{0.30}$MnO$_3$ (PCMO30) is far more complex than the simple percolation of large metallic clusters. Instead, 
the I-M transition in this system is associated with insulating ferromagnetic (FM) regions that become metallic in a first-order process. This result challenges 
our present understanding of the CMR effect and suggests that,  
if percolation plays an important role in the CMR process, it must be the percolation of insulating clusters in conjunction with an underlying first-order phase transition.

PCMO30 is an ideal system to test the mesoscopic phase separation scenario, because it has an inhomogeneous low-temperature insulating  state\cite{cox,radaelli,frontera} where ferromagnetism, antiferromagnetism, and charge ordering coexist\cite{hideki}. This insulating  state is metastable and can be converted to a metallic 
state by the application of an external magnetic\cite{hideki,tomioka} or electric field\cite{asamitsu}, high pressure\cite{moritomo}, and by exposure to x-rays\cite{kiryukhin} or visible light\cite{miyano}. 
In the mesoscopic electronic phase separation picture \cite{uehara,fath}, the electrical conductivity in the CMR materials is achieved via the percolative transport of carriers through micrometer-sized FM metallic domains in an insulating antiferromagnetic (AF) background. In this scenario, the effect of an external magnetic field in a system like PCMO30 would be only to enlarge the FM metallic clusters, at the expense of the AF insulating regions, without modifying the magnetic coupling of the spins.
We use inelastic neutron scattering to probe the microscopic magnetic coupling  associated with the FM clusters in PCMO30. We achieve this by measuring the FM spin-wave (SW) stiffness constant ``$D$''(which  measures the strength of the magnetic coupling of the spins \cite{stiffness}) as PCMO30 undergoes the I-M transition induced by a magnetic field. 
 The use of thermal neutrons, which probe length scales of the order of 100-300\AA, ensures that our measurements are insensitive to FM cluster-size changes at micron scales.  
For a conventional  ferromagnet the effect of an external magnetic field of a few Tesla is limited to induce a small Zeeman gap in the SW dispersion relation,  with no change in $D$ \cite{lovesey}. Surprisingly, we discovered that when the I-M transition in PCMO30 is induced by a 3.5 T field at low temperature there is also a three fold discontinuous change in the magnitude of $D$.  This unexpected result indicates that the I-M transition involves a first-order transformation from insulating to metallic clusters, and not just the simple percolation of metallic clusters. Thus, any realistic percolation model of the I-M transition must account for the microscopic interactions of the spins and the first-order nature of the I-M transition. 

For this study, we prepared a single crystal of PCMO30 by the floating-zone method \cite{tomioka}.  Transport and electron-probe microanalysis on different parts of the crystal  indicated that the Ca doping was homogeneous. The crystal has a  mosaic spread of about 1 degree and a volume of 0.4 cm$^3$.  The neutron scattering measurements were performed at Oak Ridge National Laboratory's High Flux Isotope Reactor. Most of the measurements were performed using triple-axis spectrometers configured to provide an energy resolution at the elastic ($E=0$) position of $\Delta E =0.55meV$ (full-width at half maximum (FWHM))\cite{instrument}.
Neutron diffraction measurements were also carried out at the Wide Angle Neutron Diffractometer (WAND)\cite{wand1,wand2}. 

PCMO30 has an orthorhombic structure slightly distorted from the cubic lattice. 
For simplicity, we  label the wave vectors $Q=(q_x,q_y,q_z)$ in units of
\AA$^{-1}$ as $(H,K,L)=(q_xa/2\pi,q_ya/2\pi,q_za/2\pi)$ in the
reciprocal lattice units (rlu) appropriate for the pseudo cubic unit cells with 
lattice parameter $a\approx 3.87$ \AA. In this convention all $(H, K, L)$ ($H, K, L$ integers) reflections are allowed. The crystal was oriented to allow wave vectors of the form $(H,K,0)$ to be accessible in the scattering plane. 
PCMO30 has a complex sequence of transitions.\cite{hideki} Below $T_{CO}\approx$ 200K it exhibits charge ordering while remaining paramagnetic, below $T_N \approx$140K  the magnetic moments associated with the Mn$^{3+}$ and Mn$^{4+}$ ions order 
antiferromagnetically in the so-called pseudo CE-type structure, and only below $T_C \approx$ 110K the system develops a FM  component that coexists with the CE-type antiferromagnetism. At zero magnetic field PCMO30 remains an insulator at all temperatures. The charge-ordered and
antiferromagnetic phases are both characterized by satellite reflections $( 1/4, 1/4, 0)$ and $(1/2, 0, 0)$ (and equivalent reflections).

We collected data under two different conditions, first by applying a magnetic field after the sample was
cooled in zero field (zero field cooling or ZFC), and  by
cooling the sample under an applied field (field cooling of FC). The magnetic field was applied perpendicular to the scattering plane by an Oxford cryomagnet (0T$\le H \le$ 7.0 T). 
The ZFC with H=0 measurements revealed the clear presence of diffuse scattering around the  main Bragg peaks above $T_C$, (see Fig. 1) . This diffusive component is strongest at the smaller wavevectors and becomes gradually weaker as the wavevector 
increases, as expected for the magnetic form factor of the Mn ions, indicating that this scattering is magnetic. When the sample is cooled below $T_C$ a long-range FM component \cite{long_range} develops (additional Bragg intensity at the $(H, K, L)$ 
positions), at the same time that the diffuse scattering becomes weaker. This diffusive component, however, does not vanish even at the lowest temperatures indicating the inhomogeneous character of this system at zero field. 
In addition to the diffuse scattering  there are additional superlattice reflections indicating the antiferromagnetism associated with the charge ordering of the pseudo CE-type. This ordering has characteristic wavevectors $(\pm 1/4, \pm 1/4, 0)$ 
(see Fig. 1). From the sharpness of the
charge ordering peaks, it is clear that the charge ordering in PCMO30 is long-range \cite{long_range} in contrast to the short-range charge ordering in La$_{1-x}$Ca$_x$MnO$_3$ family of materials \cite{dai1,adams}.
 
Figure 2 shows the ZFC elastic scattering profiles of the (1 0 0) Bragg peak at T=40K measured with a triple-axis spectrometer, for H=0, 2.5 and 3.5T. The long-range Bragg 
component\cite{long_range} of the scattering has been fitted to a Gaussian while the magnetic diffusive component has been fitted to a Lorentzian lineshape.  The magnetic diffuse scattering indicates the presence of short-range FM clusters with a correlation length  $\xi \approx$ 40 \AA. 
These clusters are static within a $\Delta t \approx$ 1ps and coexist with long-range FM clusters \cite{long_range}. The effect of the magnetic field  is to gradually reduce the number of the short-range FM clusters without significantly changing 
their correlation length, at the same time that the ferromagnetic moment is increased.  
This is reflected by the intensity reduction of the Lorentzian component 
without significant change in  its linewidth, while 
the intensity of the Bragg component increases (Fig. 2).  When the system is FC at 6.8T, there is no trace of the Lorentzian component.
The (ZFC) magnetic field dependence of the intensities of these two components has been plotted in figure 3a. 
The region where there is a jump in the intensity of the FM component (around 3.5T) corresponds to the I-M transition, which is also the region where the CE-type antiferromagnetism is greatly (but not completely) 
suppressed.\cite{hideki}  Thus, in the absence of a field there is a coexistence of a)long-range FM \cite{long_range},  b) short-range FM clusters ($\xi \approx$ 40 \AA), and c) long-range CE-type AF regions\cite{long_range}. 
The effect of the magnetic field is to gradually incorporate the short-range FM clusters into the long-range FM regions\cite{long_range} and, at the same time, to gradually suppress the CE-Type AF clusters.

We also measured the spin waves in PCMO30 both in the FC and ZFC conditions. First we measured the spin waves at 40K after the sample was cooled in a 6.8T field. Then the field was gradually decreased to 5T, 2T and 0.5T and the low-{\it q} ($q \leq$ 0.07 rlu) spin waves were measured for every field.
The measurements were made in the constant-$Q=(1+q, 0, 0)$ mode. The SW energies followed the usual quadratic dependence expected for a ferromagnet in the long-wavelength (small {\it q}) limit: $E(q)=\Delta + D q^2$, where $\Delta$ is the Zeeman energy gap and $D$ is the SW stiffness coefficient\cite{lovesey}. This is shown in figure 4a, where the measured FC low-$q$ SW energies have been plotted vs. $q^2$ for various magnetic fields, the straight lines are the result of fits to the quadratic 
dispersion relation.  The upper inset of figure 4a shows a plot of the fitted SW stiffness coefficient $D$ vs. H. Within the experimental error the SW stiffness  does not change with the field (after FC at 6.8T), and has a value of $D=(140\pm5)$meV-\AA$^2$ at T=40K. Thus,
the only effect of the field in the FC case is to open a Zeeman energy gap that grows linearly with the applied field as $\Delta =g \mu H$ (see lower inset of fig. 4a). 
We note that the magnetic moment obtained from this plot is only $\mu=(2.57 \pm 0.43)\mu_B$, which is smaller than the expected full Mn moment of 3.7$\mu_B$ for the $x=0.30$ doped Mn$^{3+}$/Mn$^{4+}$ system.
The low-{\it q} SW excitations at T=40K in the ZFC condition were different. Unlike the FC case, the ZFC measurements had to be performed in the constant-E mode due to the high background produced by the strong FM diffusive component mentioned above, 
which is strongest in the vicinity of $Q=(1, 0, 0)$.  In all cases the SW energies also followed the expected quadratic dependence, the results from these measurements have been plotted in fig. 4b. The surprising result is 
that the ZFC SW stiffness coefficient for fields up to $\approx$ 3T is only $D\approx \ $50meV-\AA$^2$, almost a factor of three smaller than for the FC case. The full stiffness $D$ = (140$\pm$5)meV-\AA$^2$ is only recovered when the I-M transition is induced at $H\approx$ 3.5T
(see Fig. 3b), at the same time when the $\xi \approx$ 40\AA $\ $clusters are suppressed.
 
The sharp three-fold change in $D$ suggests either a drastic change in  the magnitude of the spins or a drastic change in the strength of the spin interactions at the PCMO30 field-induced I-M transition. 
The first possibility can be ruled out from Fig. 3a as the magnitude of the small change in the FM component cannot account for the magnitude of the observed change in $D$.  
Our findings thus favor the drastic change in the strength of the microscopic spin interactions at the I-M transition. We note that the high SW-stiffness $D=(140\pm5)$mev-\AA$^2$ is comparable to other FM metallic manganites near the $x=0.30$ doping \cite{jaime}, where double exchange seems to be the dominant magnetic interaction. The low SW-stiffness $D \approx \ $50meV-\AA$^2$ is similar to that of the recently measured FM insulator  $La_{0.80}Ca_{0.20}$MnO$_3$\cite{dai2,jeff1}, a system likely to be dominated by some form of superexchange, and seems to be characteristic of the FM insulating manganites\cite{okuda}.

The results of our microscopic measurements on PCMO30 support a picture in which the insulating phase contains large (long-range\cite{long_range}) and small ($\xi \approx$ 40\AA $\ $) FM ``insulating" (low SW-stiffness) clusters,
with no evidence of any large FM metallic (high SW-stiffness) regions. 
The application of an external magnetic field reduces the number of the small clusters until the system becomes metallic in a first-order transition. 
This is consistent with two recent reports of thermodynamic  measurements in PCMO30. Roy {\it et al} \cite{roy} reported an enormous release of energy and a strong irreversibility at the I-M transition of this system. Deac {\it et al}\cite{deac} suggested the existence of field-induced insulating clusters in this material at low magnetic fields.    
Our results are inconsistent with the mesoscopic phase separation picture \cite{uehara,fath}, in which the I-M transition in the CMR materials is associated with the percolation of FM metallic domains, and   
show that the physics involved in this transition is far more complex.
One possible scenario of the I-M transition in the CMR materials
could involve the percolation of insulating clusters in conjuction with an underlying first-order transition. It is known that random quenched disorder may or may not produce rounding of a first-order phase transition. This was first studied by Imry and Wortis \cite{imry}, and more recently by Moreo {\it et al.}\cite{adriana}. This is a challenge to the current theoretical approaches of CMR.

Finally, we note that the nano-size FM clusters of Figure 2 are comparable in size to those observed in other CMR materials above and at their FM transition temperatures\cite{jaime,jeff,deteresa}. The field dependence of the nano-size FM clusters of figure 3a seems to imply that these nano-size clusters do play some role in the I-M transition. 

This work was supported by U.S. DOE under contract DE-AC05-00OR22725 with UT-Battelle, LLC and by JRCAT of Japan. 

\begin{figure}

 \caption{Neutron diffraction patterns of PCMO30  in the (H, K, 0) reciprocal lattice plane (pseudo-cubic notation).  The white arrows indicate the region near the $(\pm\delta,2\pm\delta,0)$ ($\delta = 1/4$) satellites 
expected for the CE-type charge ordering. The rings around the origin 
correspond to scattering from the aluminum sample holder and the narrow ``streaks''
through the strong Bragg peaks are an artifact from the detector.
}

\label{autonum}

\end{figure}

\begin{figure}

\caption{$Q=(1+q,0,0)$ scattering profiles for H=0, 2.5 and 3.5 (ZFC) and 6.8T (FC). The solid lines are the results of least-square fits to Lorentzian (diffuse) and Gaussian (Bragg) components, except for the FC data which has only a Gaussian 
component.
}

\end{figure}

\begin{figure}

\caption{3a) ZFC magnetic field dependence of the intensities of the Bragg (solid triangles) and Lorentzian (open squares) components of the $(1, 0, 0)$ FM reflection.   The broken lines are a guide to the eye. The dotted line at 
$H\approx$3.5T indicates the I-M boundary. 3b) ZFC magnetic field dependence of the low-q SW stiffness coefficient $D$ at T=40K.}

\end{figure}

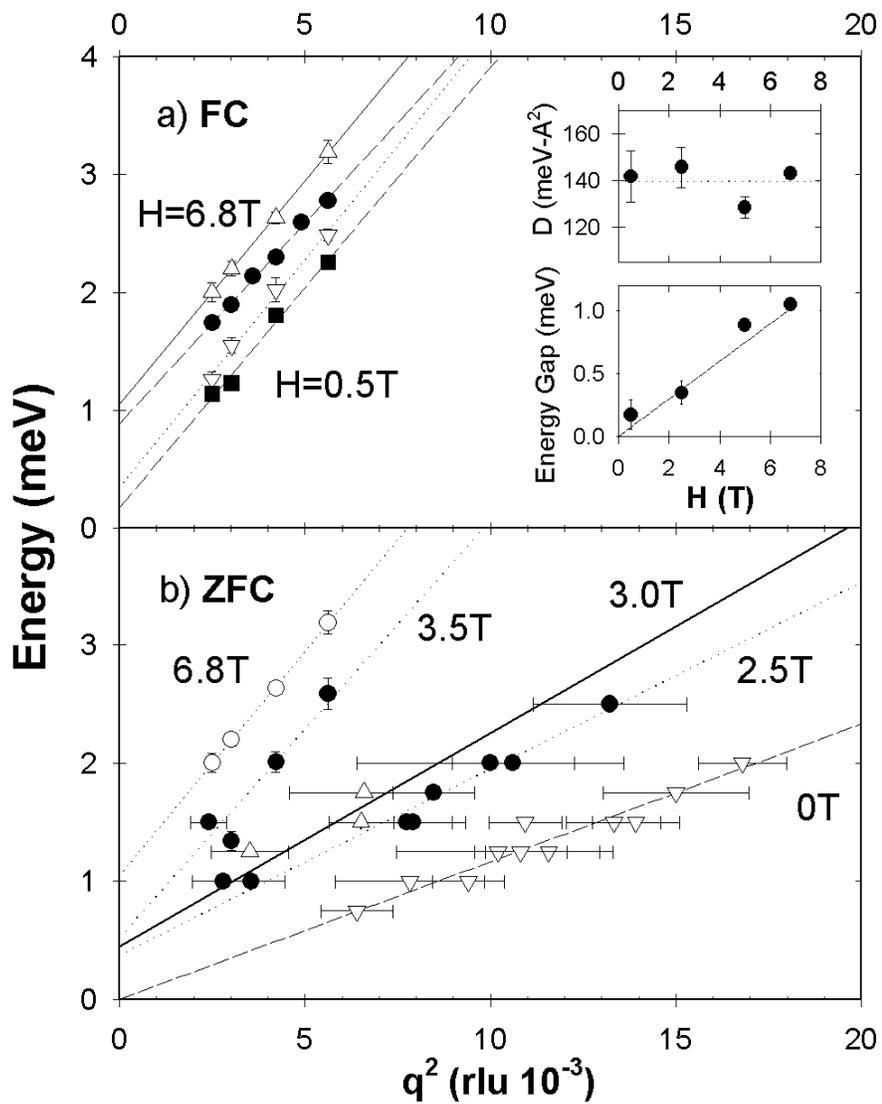
\begin{figure}
\caption{4a) FC low-q SW energies vs $q^2$ showing the usual quadratic dispersion $E=\Delta + Dq^2$ expected for a FM. Upper inset:  FC SW stiffness coefficient vs. H.
Lower inset: Fitted  energy gap $\Delta$ vs. magnetic field showing the expected linear Zeeman relation.  4b) ZFC low-q SW energies vs $q^2$ showing the usual quadratic dispersion $E=\Delta + Dq^2$ expected for a FM. Note the drastic change in 
the slope ($D$ coefficient) at $H\approx$ 3.5T, when the I-M transition occurs.
 }

\end{figure}


\begin{references}
\bibitem{stiffness} The SW stiffness constant $D$ is proportional to magnetic exchange constant $J$ in the Heisenberg ferromagnet model, and to the kinetic energy term in the DE model.
\bibitem{long_range} Within our probe ``long-range" means larger than a few hundred angstroms.
\bibitem{instrument} This was obtained using pyrolytic graphite (PG)  as monochromator (PG(0 0 2)) and filters, and a beryllium analyzer (Be(1 0 1)). The 
final neutron energy was fixed at $E_f=13.6$ meV with collimations of, proceeding from the reactor to the detector, 40-20-40-120 min (FWHM).
\end{references}
\end{document}